\documentstyle[preprint,aps,eqsecnum]{revtex}
\def\beq{\begin{equation}}
\def\eeq{\end{equation}}
\def\ra{\rightarrow}
\def\B{\overline B}

\begin{document}
\setcounter{section}{1}
\def\lest{\;\raisebox{-.4ex}{\rlap{$\sim$}} \raisebox{.4ex}{$<$}\;}
\def\gest{\;\raisebox{-.4ex}{\rlap{$\sim$}} \raisebox{.4ex}{$>$}\;}
\def\eqst{\;\raisebox{+1.5ex}{\rlap{?}} \raisebox{.01ex}{$=$}\;}
{\tighten
\preprint{\vbox{\hbox{FERMILAB--PUB--97/323--T}}}

\title{CP Asymmetries in (Semi-)Inclusive $B^0$ Decays}
\author{Isard Dunietz}
\address{Fermi National Accelerator Laboratory \\ P.O.~Box 500,
Batavia, Illinois 60510}

\bigskip
\date{\today}

\maketitle
\begin{abstract}
It was recently pointed out that inclusive $B^0(t)$ decays could show
CP violation.  The totally inclusive asymmetry is expected to be tiny
[${\cal O}(10^{-3})$] because of large cancellations among the
asymmetries in the charmless, single charm and double charm final
states. Enriching particular final state configurations could
significantly increase the CP-asymmetry and observability.  Such
studies can extract fundamental CKM (Cabibbo-Kobayashi-Maskawa)
parameters, and (perhaps) even $\Delta m(B_s)$. A superb vertex
detector could see CP violation with $10^5 \; (10^6)$ flavor-tagged
$B_s \; (B_d)$ mesons within the CKM model. Because the effects could
be significantly larger due to new physics, they should be searched
for in existing or soon available data samples.

\end{abstract} 
}
\newpage

CP violation remains a mystery more than 30 years after its
discovery~\cite{ccft}.  It has been observed so far only in $K^0$
decays. Our entire knowledge can be summarized by the single
CP-violating quantity~\cite{pdg}
\beq
\epsilon =A(K_L\ra 2\pi )/A(K_S\ra 2\pi )=2.28 \times 10^{-3} \times e^{i \pi/4}\;.
\eeq
CP violation is not just a quaint, tiny effect in $K^0$ decays, but is
necessary for baryogenesis~\cite{sakharov}. The origin of CP violation
has not yet been established. A fundamental understanding of CP
violation will bring about a deeper appreciation of our existing
universe. The fashionable CKM hypothesis \cite{ckm} allows for one
CP-violating phase which is fitted to the single observed quantity
$\epsilon$. In contrast, other aspects of the Standard Model have been
subjected to many independent tests and have been verified to high
precision~\cite{pdg}.  Fortunately, the CKM hypothesis is testable and
predicts large CP-asymmetries in many $B$ decays
\cite{fleischerreview}, for instance \cite{bigisanda}
\beq
{\rm Asym} (B_d \ra J/\psi K_S )\gest 20\%\;.
\eeq

The traditional efforts focused on the gold-plated $B_d\ra J/\psi K_S$
or other exclusive $B$-modes. While the CP-asymmetry is predicted to
be large, the effective branching ratio is tiny ($\sim 10^{-5}$).
Orders of magnitude larger branching ratios are available from studies
of (semi-) inclusive CP-asymmetries
\cite{sachsd,jarlskog,bbd,stodolsky,b20,grossman},
\beq
\label{incl}
I(t) \equiv \frac{\Gamma (B^0 (t)\ra all) -\Gamma (\B^0 (t)\ra
all)}{\Gamma (B^0 (t)\ra all) +\Gamma (\B^0 (t)\ra all)} \;.
\eeq

Such an inclusive asymmetry appears to violate the CPT theorem, which
guarantees equal total widths for particle and antiparticle. This
theorem may have discouraged experimenters to search for CP effects in
their large, inclusive $B$-samples.  There is no contradiction with
the CPT theorem, however. $B^0-\B^0$ mixing introduces an additional
amplitude, which permits the time-dependent totally inclusive rate to
differ from its CP-conjugated partner. The only constraint provided by
the CPT theorem is that
\beq
\int^{\infty}_0 dt\; \Gamma (B^0 (t)\ra all) =\int^{\infty}_0 dt\; \Gamma (\B^0 (t)\ra all)\;.
\eeq
For a \underline{truly} unbiased $\stackrel{(-)}{B^0}$ sample, the
time-dependence is known \cite{bbd,b20},
\beq
\label{unbiasedasym}
I(t) =a\left[\frac{x}{2}\sin \Delta mt-\sin^2 \left(\frac{\Delta
mt}{2}\right)\right]\;.
\eeq
Here the mixing parameter $x\equiv \Delta m/\Gamma$. The
width-difference $\Delta\Gamma$ is neglected throughout this report,
and $a$ is the conventional dilepton asymmetry~\cite{dilepton},
\beq
\label{dilepton}
a \equiv Im (\Gamma_{12}/M_{12})= - \frac{\Gamma (B^0 (t)\ra W)-\Gamma
(\B^0 (t)\ra \overline W)}{\Gamma (B^0 (t)\ra W) +\Gamma (\B^0 (t)\ra
\overline W)}\;.
\eeq
Here $W$ stands for a flavor-specific $\overline B^0$ mode, i.e. that
cannot be accessed from an unmixed $B^0$, such as $\ell^- X$.

The observable $a$ is expected to be tiny $[\sim 10^{-3} \; (\lest
10^{-4})$ for $B_d \; (B_s)$ mesons]~\cite{lusignoli,buchalla}.  Much
larger CP violating effects are expected in each of the semi-inclusive
$b\ra c\!\!/$ (charmless), $\stackrel{(-)}{c}$ (single charm),
$c\overline c$ (double charm) transitions~\cite{bbd}.  The
semi-inclusive asymmetries are opposite in sign and largely cancel
when combined to form the totally inclusive asymmetry $a$.  A superb
vertex detector could select each of the semi-inclusive transitions,
thereby becoming sensitive to CP violating effects that are predicted
to be significantly enhanced.  The selection could be done
continuously by varying the efficiencies $\epsilon_i$ for recording
the specific transitions (see Table 1).  The efficiencies to observe
charmless, single charm, double charm final states are denoted by
$\epsilon_0$, $\epsilon_1$, $\epsilon_2$, respectively.  Because
vertexing alone cannot distinguish $B^0$ modes involving hidden
charmonia from truly non-charm final states, both are classified as
charmless modes in this note.

This report assumes identical detection efficiency for mode
$\epsilon_i$ and CP-conjugated mode $\overline\epsilon_i$,
\beq
\epsilon_i =\overline\epsilon_i \;.
\eeq
The assumption may not hold because the detector is made out of matter
and because of possible asymmetries in reconstructing positive versus
negative tracks.  Since those are detector-specific issues, they will
not be considered further in the main text (see, however,
Appendix~\ref{faking}), but have to be investigated by each
experiment.

The efficiencies can be varied continuously by suitable cuts, thereby
``biasing" or ``weighting" the inclusive asymmetry Eq.~(\ref{incl})
and making it dependent on $\epsilon_i$,
\beq
\label{weightincl}
I(t) =-a\;\sin^2 \left(\frac{\Delta mt}{2}\right) +c \;\sin\Delta
mt\;.
\eeq
Here $a$ is the dilepton asymmetry defined in Eq.~(\ref{dilepton}) and
is independent of $\epsilon_i$, while the coefficient $c$ depends on
$\epsilon_i$. Both coefficients $a$ and $c$ are functions of CKM
parameters and are given in Appendix~\ref{ckmapp}.\footnote{It is now
clear how to extract the efficiency-independent observable $a$ from
time-dependent and efficiency-varying studies. The extraction can be
accomplished even for a non-vanishing width difference
$\Delta\Gamma$. The formalism is straightforward, just somewhat more
cumbersome~\cite{bbd}.}  Alternatively, one could assign to each
inclusive $B^0 / \overline B^0$ decay a probability for being a
charmless, single charm or double charm transition, thereby
``weighting" the inclusive asymmetry.  The coefficient $a$ is
independent on this ``weighting", while the coefficient $c$ depends on
it.

For identical detection efficiencies $\epsilon_i =\epsilon$,
Appendix~\ref{ckmapp} obtains $c=a\cdot x/2$ and the truly inclusive
asymmetry is recovered. Further note that in general a time-integrated
CP violating asymmetry survives, since $c$ normally differs from
$a\cdot x/2$. This realization permits us to search for
\underline{time-integrated} CP violating effects in single or double charm or charmless samples.

Our current knowledge about the CKM matrix in the Wolfenstein
representation~\cite{wolfenstein} can be parameterized as follows
\cite{buchalla,nir} $$-0.3 < \rho < 0.3,$$ $$0.2 <\eta < 0.5 .$$ The
effect on $c$ of varying $\rho$ is not too significant, whereas
varying $\eta$ has a more drastic effect (see Appendix~\ref{ckmapp}).

Choose $\rho=0$ and $\eta =0.4$ for illustrative purposes. As a
function of efficiencies $\epsilon_i$, Tables II and III list the
CP-violating coefficient $c$. The last column shows how many tagged
$B^0\; [N_{B^0}]$ and tagged $\B^0\; [N_{\overline B^0}]$ have to be
produced to observe $c$ to $3\sigma$ accuracy (with $a$ neglected).
Here tagging denotes the distinction of an initial $B^0$ and
$\overline B^0$.  A superb vertex detector could observe inclusive
CP-violation with $10^5\; (10^6)$ tagged $B_s\;(B_d)$ mesons. Because
the specific efficiencies $\epsilon_i$ can be varied continuously,
many systematic effects can be controlled and studied. For a given
detector, the optimal choice for $\epsilon_0,\epsilon_1,\epsilon_2$
can be determined, by minimizing the required production of tagged
$B^0$ and tagged $\B^0$ mesons to observe a $3\sigma$ asymmetry
$[N_{B^0} +N_{\B^0}](3\sigma)$.

A nonzero coefficient $c(B_s) \neq 0$ ($a(B_s) \neq 0$), would prove
CP violation in the $B_s$ sector and further would permit an
unconventional determination of $\Delta m(B_s)$ (from
flavor-nonspecific final states).  In contrast, conventional methods
require the $\overline B_s$ to be seen in flavor-specific modes, such
as $D^+_s X \ell^- \overline\nu, D^+_s (\pi ,\rho
,a_1)^-$~\cite{snowmass,lep}.

The double charm $\overline B_d$ modes are promising, and have a
predicted semi-inclusive asymmetry of ${\cal O} (1\%)$ (see Table
III). The CP signal is due to the Cabibbo suppressed $b\rightarrow
c\overline cd$ transitions \cite{bbd}, and is unfortunately diluted by
the $\sim 20$ times larger Cabibbo-allowed $b\rightarrow c\overline
cs$ processes. The generic $\B_d$ decays governed by $b\ra c\overline
cs$ give rise to flavor-specific final states which cannot be reached
from both an unmixed $B^0$ and an unmixed $\B^0$, and therefore are
not sensitive to the mixing-induced CP violating effects discussed in
this note.

One can either attempt to enrich the $b\ra c\overline cd$ transitions
over the $b\ra c\overline cs$ processes via particle identification,
or one could cause the modes governed by $b\ra c\overline cs$ to be
accessible from both a $B^0$ and a $\B^0$.  The latter can be
accomplished by having the primary $s$ quark hadronize into a neutral
kaon, which is then observed as a $K_S$ or $K_L$.\footnote{The CP
effects involving primary $K_S$ versus primary $K_L$ are opposite in
sign, and therefore should be combined carefully.}  More generally,
$\B^0$ modes that involve a single primary $s$ quark\footnote{The
$s$-quark produced in $\B_d$-transitions governed by $b\ra c\overline
cs, c\overline us, u\overline cs,s,$ or the spectator $s$-quark in
$B_s-$transitions governed by $\overline b\ra \overline cc\overline
d,\overline cu\overline d,\overline uc\overline d,\overline d.$} are
normally flavor-specific. Nevertheless, mixing-induced CP violating
effects are expected when that primary $s$ quark is seen as a neutral
$K_S$ or $K_L$, as in the following $B^0$ modes:
\beq
\label{ksc}
{\rm primary}\; K_{S(L)}+\left\{\not\!c ,\stackrel{(-)}{c}, c\overline
c\right\}.
\eeq

CP violation may be seen in inclusive $K_S$ studies [either
time-integrated or time-dependent]:
\beq
\label{cpks}
\frac{\Gamma (B_d (t)\ra K_S X)-\Gamma (\B_d (t)\ra K_S X)}{\Gamma (B_d 
(t)\ra K_S X)+\Gamma (\B_d (t)\ra K_SX)} \;.
\eeq

Focusing on primary $K_S$'s$\; (B\ra K_S)$, which do not originate
from intermediate charmed hadrons $(B\ra\stackrel{(-)}{c}\ra K_S)$,
may enhance the CP asymmetry [Eq.~(\ref{cpks})] within the CKM
model. The underlying transitions are essentially:
\begin{itemize}
\item $\overline B_d \ra D\overline DK_S X$ \cite{bigisanda,jarlskog}
\item $\overline B_d\ra K_S X$ governed by penguin amplitudes, and
\item $\overline B_d\ra (c\overline c)K_S X$, where the $(c\overline c)$
pair annihilates nonperturbatively into light hadrons
\cite{disy,hawaii,close} or hadronizes as hidden
charmonia~\cite{jarlskog}.
\end{itemize}
These processes are governed essentially by the CKM combination
$V_{cb}V^*_{cs}$. $B_d-\overline B_d$ mixing introduces the
interfering amplitude $B_d(t) \to \overline B_d \to K_S X$, and CP
violation could occur.  Within the CKM model, that CP violating effect
depends on the weak phase $2\beta$.  In addition to the observation of
the primary $K_S$, other available information concerning the decay
products of the $B_d/ \overline B_d$ should be incorporated, as that
may increase further the CP violating effects.

While obviously very useful, superb vertexing is not mandatory for
several studies advocated here.  For instance, the time-integrated
$B_d \to K_S X$ asymmetry (\ref{cpks}) does not require superb vertex
information. In addition, present techniques can enhance the double
charm content by fully reconstructing one charmed hadron and inferring
inclusively (via the soft charged pion in $D^*\ra \pi^+ D$ processes,
and/or via vertexing) the other in the same $b$-hemisphere. The
existence of two charmed hadrons in the same $b$-hemisphere could be
inferred more inclusively perhaps by combining vertex information with
observed kaon yields. In contrast, charmed hadrons produced in single
charm events differ in their momentum distribution and are likely more
detached from the remainder of the $b$-decay than double charm
events. Those and other available techniques could be used to enhance
CP effects in existing or soon available data samples.

This note focuses on mixing-induced CP violation which requires
tagging. Semi-inclusive $B$ decays could show direct CP violation,
which does not involve mixing-induced amplitudes and requires no
tagging \cite{direct,browdercp}. The direct CP violating effects are
expected to be tiny. If they are observed in charged $B^\pm$ decays,
then those $B^\pm$ measurements can be incorporated straightforwardly
into the general formalism of semi-inclusive $\stackrel{(-)}{B^0}$
asymmetries \cite{bbd}.  In addition to searching for quasi-inclusive
direct CP violation involving charged
\underline{primary} $K^{(*)-}$ \cite{browdercp}, mixing-induced CP effects could be looked for in
tagged momentum-spectra of \underline{secondary} $K^{(*)}\; [B^0 \ra
\stackrel{(-)}{c}\ra K^{(*)}]$~\cite{grossman,playfer}. The single
charm/double charm content can be varied somewhat by varying the
$K^{(*)}$ momenta.

What is the current experimental status? The DELPHI and SLD
collaborations~\cite{delphi,sld} implicitly assumed an unbiased
inclusive $B$ sample. By fitting their data to the known unique
time-dependence Eq.~(\ref{unbiasedasym}), they extracted the
observable $a$ for the $B_d$ meson
\beq
a_d=\left\{\begin{array}{ll}-0.022 \pm 0.030 \pm 0.011 & {\rm DELPHI}
\\ -0.04 \pm 0.12 \pm 0.05 & {\rm SLD} \; .
\end{array}\right.
\eeq
Their data samples are probably biased, however (see
Appendix~\ref{bias}).  Because in addition, the predicted $a$ is
tiny~\cite{lusignoli,buchalla} it is instructive to fit the
measurements to $\sin\Delta mt$.\footnote{Current
data~\cite{delphi,sld} are incapable of discriminating among a wide
variety of possible interpretations. On the other hand, the existing
data do not rule out a $\sin \Delta m t$-dependence. We performed a
single parameter fit to the DELPHI data~\cite{delphi} of the form $$
I(t) =c\sin\Delta mt, $$ with $\Delta m = 0.474\;$ ps$^{-1}$
\cite{pdg}, and obtained $$ c=0.03 \pm 0.01\;.  $$ Our two parameter
fit for $c$ and $\Delta m$ yields $$ c= 0.03 \pm 0.01,\;\;\;\Delta
m=0.5 \pm 0.1\;{\rm ps}^{-1}, $$ which correctly recovers the known
$B_d-\overline B_d$ frequency $\Delta m$. The fits have a $\chi^2$ per
degree of freedom somewhat better than 1. The quoted errors are
statistical only. Systematic uncertainties could be significantly
larger.}  If CP is conserved, the inclusive, time-dependent asymmetry
vanishes and cannot show any $\Delta m t-$dependence.  In real life,
however, a residual $\Delta m t-$dependence may be seen even in the
absence of CP violation, because, for example, of different detection
efficiencies for mode and CP-conjugated mode (see
Appendix~\ref{faking}).

Such "fake" CP-effects are less important when the expected CP
violating signal is enhanced manyfold.  The enhancement can be
accomplished by refined CP studies that consciously enrich specific
non-leptonic transitions. While this note discussed enrichments of the
charmless, single charm, and double charm sectors, the idea is clearly
much more general.  As more insights into $B$ decays are gained,
suitable cuts or weighting factors can be designed for each of the
sectors to further enhance CP violation, for instance, by increasing
CP-even over CP-odd configurations (or vice versa). Those enrichment
techniques are in their early stages. Once they mature, observation of
CP violation and quantitative extractions of CKM parameters become
feasible \cite{bbd}.  That may prove useful, because CP violation is
one of the most important mysteries in high energy physics.

\section{Acknowledgements}

We are grateful to C. Kreuter for informing us about the existence of
Ref.~\cite{delphi}, and for enlightening discussions.  We thank the
1997 DELPHI spokesperson D. Treille for permitting us to use DELPHI's
public data, C. Kreuter for mailing us the numerical values,
G.~Buchalla, M. Daoudi and Su Dong for conversations concerning
unbiased inclusive $B_d$ samples, and R.N. Cahn, C. Gay, J. Incandela
and J.L. Rosner for important comments on an earlier draft. This work
was supported in part by the Department of Energy, Contract No.
DE-AC02-76CH03000.

\appendix

\section{On faking CP Violation}
\label{faking}
Because current $B$ decay simulations may have to be modified (see
Appendix~\ref{bias}), slight differences in acceptance and detection
efficiencies of mode $f$ and CP-conjugated mode $\overline f \equiv
CP\;f$ must be investigated further. Those differences arise because
the detector is made out of matter where particle and antiparticle
interact differently and because of possible asymmetries in
reconstructing positive versus negative tracks.  The differences are
parameterized by the small deviation from 1 of the real quantity
$\eta$ in this appendix. This appendix assumes CP conservation
throughout. A $\Delta mt$-dependence may still be seen, because
\begin{eqnarray}
\label{fake}
& &\frac{\Gamma (B^0(t) \ra f)+\eta \; \Gamma (B^0(t) \ra \overline f)
-\left\{\Gamma
\left(\overline B^0\left (t\right)\ra f\right) + \eta \; \Gamma\left(\overline
B^0\left(t\right)\ra \overline f\right)\right\}}{\Gamma (B^0(t) \ra f)
+\eta \; \Gamma (B^0 (t) \ra \overline f) + \Gamma (\overline B^0
(t)\ra f) +\eta \; \Gamma (\overline B^0 (t)
\ra \overline f)} =\nonumber \\
& & \nonumber \\ & & (1-\eta \; ) \frac{\left[\cos\Delta mt
\left(1-|\lambda |^2\right) -2 Im\lambda \sin\Delta mt\right]}{(1+\eta \;  )
\;\;(1+|\lambda |^2 )} \; .
\end{eqnarray}

The coefficients $q$ and $p$ relate the $B^0$ and $\overline B^0$
states to the mass eigenstates and satisfy $|q/p|=1$
\cite{bsbsbar}. The interference terms
$\lambda\equiv~q\;\langle~f|\overline{B}^0\rangle\:/\;(\;p\;\langle~f|B^0\rangle\;)$
and $\overline{\lambda}\equiv~p\;\langle\overline{f}|B^0
\rangle\;/\;(\;q\;\langle\overline{f}|\overline{B}^0\rangle\;)$
satisfy $\lambda =\overline\lambda$ under the assumption of CP
conservation. They could have a nonzero imaginary part only due to a
final state phase difference \cite{drosner,bsbsbar}. (CP conservation
demands vanishing weak phase differences!)  Eq.~(\ref{fake}) can be
traced back to the fact that a $\Delta mt$-dependence survives if the
difference between $f$ and $\overline f$ has not been accounted for
correctly $(\eta \; \neq 1)$:
\begin{eqnarray}
\Gamma (B^0 (t) \ra f) & + & \eta \; \Gamma (B^0 (t) \ra \overline f)  = \Gamma (B^0\ra f)
\frac{e^{-\Gamma t}}{2} \nonumber \\
& \times & \Big\{ (1+\eta ) (1+|\lambda |^2) + (1-\eta
)\left[\cos\Delta mt\left(1-|\lambda |^2\right)-2 Im\lambda\sin\Delta
mt\right]\Big\} \;,
\end{eqnarray}

\begin{eqnarray}
\Gamma (\overline B^0 (t) \ra f) & + & \eta \; \Gamma (\overline B^0 (t) \ra \overline f)
 = \Gamma (B^0 \ra f) \frac{e^{-\Gamma t}}{2}
\nonumber \\
& \times & \Big\{ (1+\eta )(1+|\lambda |^2) - (1-\eta ) \left[\cos
\Delta mt\left(1-|\lambda |^2 \right)-2 Im\lambda\sin\Delta
mt\right]\Big \} \;.
\end{eqnarray}
All $\Delta mt$-dependence is gone when mode and CP-mode are summed
over ``properly" $(\eta =1)$:
\begin{eqnarray}
\label{no_dmt}
\Gamma (B^0 (t) \ra f) +\Gamma (B^0 (t) \ra\overline f) & = & \Gamma (\overline B^0 (t) \ra
\overline f) +\Gamma (\overline B^0 (t) \ra f) \nonumber \\
& = & \Gamma (B^0 \ra f) e^{-\Gamma t} (1+|\lambda|^2).
\end{eqnarray}
Thus, as long as mode and CP-mode are combined properly, no $\Delta
mt$-dependence survives (\ref{no_dmt}). This is true whether or not
there exists an unaccounted difference in tagging (distinguishing) an
initial $B^0$ and $\overline B^0$.

There exist methods that may reduce a possible small discrepancy of
distinguishing an initial $B^0$ and $\overline B^0$ (for instance, by
using polarized $Z^0$'s~\cite{atwood}).  Nonetheless, we wish to
present the expression which takes that discrepancy also into
account. [The small deviation from 1 of the parameter $\tau$
quantifies the discrepancy here]:
\begin{eqnarray}
& & \frac{ \Gamma (B^0(t)\ra f)+\eta \; \Gamma (B^0 (t) \ra \overline
f) -\tau \left[\Gamma
\left(\overline B^0 (t)\ra f\right) +\eta \; \Gamma \left(\overline B^0 (t) \ra \overline
f\right)\right]} {\Gamma (B^0(t)\ra f)+\eta \; \Gamma (B^0 (t) \ra
\overline f) +\tau \left[\Gamma
\left(\overline B^0 (t)\ra f\right) +\eta \; \Gamma \left(\overline B^0 (t) \ra \overline
f\right)\right]} = \nonumber \\ & & \nonumber \\ & & =\frac{(1+\eta \;
)(1-\tau )(1+|\lambda |^2) + (1-\eta \; ) (1+\tau) \left[\cos\Delta mt
\left(1-|\lambda |^2\right) -2Im\lambda\sin\Delta mt\right]} {(1+\eta
\; )(1+\tau )(1+|\lambda |^2) + (1-\eta \; ) (1-\tau) \left[\cos\Delta
mt \left(1-|\lambda |^2\right) -2Im\lambda\sin\Delta mt\right]}
\approx \nonumber
\\
& & \nonumber \\ & & \approx\frac{(1-\tau )(1+|\lambda |^2 ) +(1-\eta
\; ) \left[\cos\Delta mt\left(1-|\lambda |^2 \right) -2
Im\lambda\sin\Delta mt\right]} {2( 1 +|\lambda |^2 )}\;.
\end{eqnarray}
That concludes our discussion of some of the systematic effects that
are CP conserving.

\section{The (Semi-)Inclusive CP Violating Parameters}
\label{ckmapp}

The time-dependence of the (semi-) inclusive CP violating asymmetry,
\footnote{Neglecting $\Delta\Gamma$ and direct CP violation.}
\beq
I(t)=-a\sin^2\left(\frac{\Delta mt}{2}\right) +c\sin\Delta mt\;,
\eeq
follows from the formalism outlined in Refs. \cite{bbd,bbdBs}.  The
coefficient $a \equiv Im(\Gamma_{12}/M_{12})$ does not depend on the
efficiencies $\epsilon_i$ but does depend on the CKM
parameters~\cite{dilepton}.  In contrast, the parameter $c$ depends
both on $\epsilon_i$ and on CKM parameters,
\beq
c=\frac{x}{2} \;\frac{\sum_{f = 0,1,2} \epsilon_f
Im(\Gamma_{f,12}/M_{12})}{(\epsilon_0 B_0 +\epsilon_1 B_1 +\epsilon_2
B_2)} \;.
\eeq
Here $B_i\; (i=0,1,2)$ denote the inclusive branching ratios (for
$\not\! c , \stackrel{(-)}{c}, c\overline c$ modes of an unmixed
$B_{d,s}$), and are listed in Table I.  Note that $c=\frac{x}{2} a$
for $\epsilon_0 = \epsilon_1 = \epsilon_2$ because $\Gamma_{12}
=\sum_{f=0,1,2}\Gamma_{f,12}$ \cite{bbd}.  What remains is to show how
$\sum_f \epsilon_f Im(\Gamma_{f,12}/M_{12})$ depends on the
fundamental CKM parameters and on other quantities.

We consider two scenarios for $B_{d,s}$ modes containing a $c$ quark
and a $\overline c$ quark. Theory estimates the inclusive CP asymmetry
for such modes \cite{bbd}.  Those modes consist of (open $c\; +$ open
$\overline c$) subchannels and (hidden $c\overline c)$
subchannels. Scenario A assumes that both subchannels experience the
same CP asymmetry, which therefore is taken to be the ``calculated"
($c$ quark $+ \overline c$ quark) asymmetry.

On the other hand, perturbative QCD favors a much suppressed asymmetry
for the (hidden $c\overline c$)
subchannels~\cite{benekebuchallathank}. Scenario B assumes that the
entire calculated ($c$ quark $+\overline c$ quark) asymmetry resides
in the (open $c+$ open $\overline c$) subchannels, with no asymmetry
in (hidden $c\overline c$) processes.

The above distinction is important because the truly (no charm) and
the (hidden $c\overline c$) modes both involve a single $B$ decay
vertex, which the main text denotes as charmless modes. Note further
that the main text denotes the (open $c +$ open $\overline c$)
channels as $c\overline c$.  The formalism yields:
\begin{eqnarray}
\sum_f\epsilon_f Im (\Gamma_{f,12}/M_{12}) = & - &\frac{\pi}{2}\;\frac{m^2_b}{M^2_W
\eta_B S_0 (x_t)}
\;\left\{Im\left(\frac{\lambda_c}{\lambda_t}\right)^2\left[\frac{\epsilon_2
B_2+\epsilon_h B_h}{(B_2+B_h)}F_2-2\epsilon_1F_1
+F_0\epsilon_0\right]\right.+
\nonumber \\
& - & \left. 2Im\frac{\lambda_c}{\lambda_t}\left[\epsilon_1 F_1
-\epsilon_0 F_0
\right]\right\} \;\;[{\rm scenario} \; {\rm A}]  \\
= & - & \frac{\pi}{2} \;\frac{m^2_b}{M^2_W \eta_B S_0 (x_t )}\;\left\{
Im\left(\frac{\lambda_c}{\lambda_t}\right)^2 \left[\epsilon_2
F_2-2\epsilon_1 F_1 +\epsilon_0 F_0 \right] + \right. \nonumber \\ & -
& \left. 2 Im \frac{\lambda_c}{\lambda_t} \left[\epsilon_1 F_1
-\epsilon_0 F_0
\right]\right\}\;\;[{\rm scenario} \; {\rm B}].
\end{eqnarray}
The QCD parameter $\eta_B = 0.8475$ and the $S_0 (x_t) =2.41$ function
dependent on $x_t\equiv (m_t /M_W)^2$ are reviewed in Ref.~\cite{bbl}.
The inclusive branching ratio [detection efficiency] into (hidden
$c\overline c$) modes is denoted by $B_h\; [\epsilon_h]$.  This report
assumes $\epsilon_h = \epsilon_0$. Because $B_{{\rm no}\; {\rm charm}}
\approx 0.01$ and $B_0$ has a predicted central value of
$0.07$~\cite{baryon97}, we chose $B_h = 0.06$ for illustrative
purposes.  Table IV lists the relevant CKM combinations $[\lambda_k
\equiv V^*_{kd} V_{kb}\; (V^*_{ks} V_{kb})$ for $\B_d\; (\B_s)$
mesons] in terms of the Wolfenstein parameters.  The $F_i\; (i=0,1,2)$
are QCD corrected phase-space factors. Their leading order expressions
in $1/m_b$ expansion are \cite{bbdBs}
\begin{eqnarray}
F_2 = \frac{\sqrt{1-4z}}{3} & & \left\{ 4\left[2\left(1-z\right)K_1
+\left(1-4z\right)K_2 \right]\right. + \nonumber \\ & &
\left. 5\left(1+2z\right)\left(K_2-K_1\right) \right\}\;, \\ F_1 =
\frac{(1-z)^2}{3} & & \left\{4\left[\left(2+z\right)K_1
+\left(1-z\right)K_2\right]\right. + \nonumber \\ & &
\left.5\left(1+2z\right)\left(K_2-K_1\right) \right\}\;, \\ F_0 =
\frac{1}{3} & & \left\{ 4\left(2K_1 +K_2 \right)
+5\left(K_2-K_1\right)\right\} \;,
\end{eqnarray}
where $$z\equiv m^2_c/m^2_b \;.$$ The QCD coefficients were taken to
be $K_1 =-0.3876$ and $K_2 =1.2544$. In addition, the numerical
estimates of Tables II--III used $m_b=4.8$ GeV/$c^2$ and $m_c=1.4$
GeV/$c^2$.

\section{Inclusive $b$-Hadron Decays}
\label{bias}

Inclusive $B$ decays maybe more subtle than currently modeled. Thus,
what is considered an unbiased inclusive $B$ data sample may in
reality be biased.  This appendix questions the current modeling of
sizable fractions of $B$ decays, especially:
\begin{enumerate}
\item baryon production in $B$ meson decays,
\item $\overline B\ra D\overline D\;\overline KX$ transitions,
\item $\overline B\ra$ no open charm, and
\item $b\ra c\overline ud$ transitions.
\end{enumerate}

\subsection{$B\ra$ baryons}

Models conventionally assume that a weakly decaying charmed baryon is
produced in generic $B\ra$ baryons
transitions~\cite{baryon92}. However, a straightforward analysis
predicts that $\overline B\ra DN\overline N' X$ processes may be a
sizable fraction of all $B\ra $ baryons transitions, where $N^{(')}$
denotes a nucleon~\cite{recal,baryon97}.  While the $\Xi_c$ yield in
$B$ decays had been neglected initially
\cite{baryon92}, its current central value \cite{browder} is too high, as can be
inferred from the more accurately measured $B\ra
\stackrel{(-)}{\Lambda_c}$ yields~\cite{recal,baryon97}. Further, the
true $\Lambda_c$ yields in $\overline B$ decays is predicted to be
reduced significantly from presently accepted
values~\cite{recal,baryon97}.

\subsection{$\overline B\ra D\overline D \;\overline KX$}
Refs.~\cite{bsbsbar,bdy} predicted a sizable wrong charm $\overline
D\;(\equiv \overline D^0, D^-)$ yield in $b$-decays, which has been
confirmed later by CLEO \cite{thorndike,briere}, ALEPH \cite{barate}
and DELPHI~\cite{delphiDbar,bplus,bsgdelphi}.  These processes were
left out in the simulations of DELPHI and SLD, thereby introducing a
bias in the supposedly totally inclusive $B$ decays.

Once the current $b\ra \overline D$ measurements are incorporated,
large uncertainties still remain.  The $B(b\ra\overline D )$ is poorly
measured at present, and so is the fraction of the time the wrong sign
$\overline D$ is seen as a $D^-$ versus $\overline D^0$, which is
important for the simulation because of differences in lifetimes and
decay patterns.  Future studies of $b\ra \overline D$ and $b\ra
D^{*-}$ will shed light on those issues~\cite{recal,isitalk97}.

\subsection{$b\ra$ no open charm}

The recent flavor specific $b\ra \overline D$ measurements made it
possible to predict $B(B\ra$ no open charm) in a variety of ways
\cite{disy,hawaii}.  Either $B(B\ra$ no open charm) is enhanced over
conventional estimates and about (10-20)\% \cite{kagan,disy}, or
$B(D^0\ra K^- \pi^+)$ is sizably below presently accepted values
\cite{recal,distw,montreal}, or both.  (If any turns out to be true,
current simulations of heavy flavor decays will have to be modified.)
Recent studies of DELPHI \cite{bsgdelphi} and CLEO \cite{thorndike}
appear not to support a large charmless yield in $B$ decays. In
contrast, a new SLD analysis uses all available distinguishing
characteristics to determine $B(b \to s g)$, and is consistent with a
significantly enhanced charmless yield~\cite{sld}.  The CLEO analysis
suggests a smaller $B(D^0 \to K^- \pi^+)$~\cite{baryon97}.

\subsection{$b\ra c\overline ud$}

About half of all $B$ meson decays are governed by the $b\ra
c\overline ud$ transitions. Only (10-15)\% of the $b\ra c\overline ud$
processes have been measured~\cite{browder}. The rest has to be
modeled. The current simulation essentially treats the $c$ and
spectator antiquark as one string and the $\overline ud$ as another,
and fragments the strings independently. We expect to achieve a
significant improvement in the simulation if we hadronize the
$\overline ud$ pair with low invariant mass into resonances as
observed in $\tau\ra\nu +\overline ud$ decays, and apply HQET methods
to the $b\ra c$ transition~\cite{hawaii,future}.  For $m_{\overline
ud}>m_\tau$, nonperturbative effects may become important and may be
difficult to model. The small color-suppressed amplitude is also
harder to model.

\newpage
\begin{table}
\caption{Branching ratios and efficiencies as a function of charm content in
inclusive $\B^0$ decays} \begin{tabular}{|l|c|c|} Process & Branching
 Ratio & Efficiency \\
\hline
\hline
$b\ra \;\not\! c$ (charmless) & 0.07 & $\epsilon_0$ \\
\hline
$b\ra \stackrel{(-)}{c}$ (single charm) & 0.74 & $\epsilon_1$ \\
\hline
$b\ra c\bar c$ (double charm) & 0.19 & $\epsilon_2$ \\
\end{tabular}
\end{table}

\begin{table}
\caption{The coefficient $c$ and required number of tagged $B_s+\B_s$ mesons to
observe a $3\sigma$ CP violating effect as a function of the
efficiencies $\epsilon_i$. The $B_s-\B_s$ mixing parameter was chosen
as $x_s =30$, the CKM parameters as $\rho =0,\eta =0.4,$ and the CP
violating parameter $a$ was neglected. The values inside the curly
parentheses assume that the double charm asymmetry is the same for the
(hidden $c \overline c$) sector and for the (open $c$ + open
$\overline c$) channels. The values in front of the curly parentheses
assume that the entire double charm ($c$ quark + $\overline c$
antiquark) asymmetry resides in (open $c$ + open $\overline c$)
channels, and that there is no asymmetry in the (hidden $c\overline
c$) sector. (See Appendix~\protect\ref{ckmapp} for details.)}

\begin{tabular}{|l|c|c|c|c|}
$\epsilon_0$ & $\epsilon_1$ & $\epsilon_2$ & $c$ & $[N_{B_s}
+N_{\B_s}] (3\sigma)$ \\
\hline
1 & 0 & 0 & $0\;\;\{-0.015\}$ & $\infty\;\; \{1\times 10^6 \}$ \\
\hline
0 & 1 & 0 & 0.007 & $6\times 10^5$ \\
\hline
0 & 0 & 1 & $ -0.023\;\; \{-0.017\}$ & $2\times 10^5\;\; \{3\times
10^5 \}$ \\
\end{tabular}
\end{table}
\begin{table}
\caption{The coefficient $c$ and required number of tagged $B_d +\B_d$ mesons to
observe a $3\sigma$ CP violating effect as a function of the
efficiencies $\epsilon_i$. The CKM parameters were chosen as $\rho =0,
\eta =0.4$, and the CP violating parameter $a$ was neglected. The
values inside the curly parentheses assume that the double charm
asymmetry is the same for the (hidden $c \overline c$) sector and for
the (open $c$ + open $\overline c$) channels. The values in front of
the curly parentheses assume that the entire double charm ($c$ quark +
$\overline c$ antiquark) asymmetry resides in (open $c$ + open
$\overline c$) channels, and that there is no asymmetry in the (hidden
$c\overline c$) sector. (See Appendix~\protect\ref{ckmapp} for
details.)}

\begin{tabular}{|l|c|c|c|c|}
$\epsilon_0$ & $\epsilon_1$ & $\epsilon_2$ & $c$ & $[N_{B_d}
+N_{\B_d}](3\sigma )$ \\
\hline
1 & 0 & 0 & $-0.005\;\; \{0.0010\}$ & $2\times 10^7\;\; \{4\times 10^8
\}$ \\
\hline
0 & 1 & 0 & $-0.0021$ & $8\times 10^6$ \\
\hline
0 & 0 & 1 & $0.009\;\; \{0.007\}$ & $2\times 10^6\;\; \{3\times 10^6
\}$
\end{tabular}
\end{table}

\begin{table}
\caption{Relevant CKM combinations in terms of the Wolfenstein parameters $(\eta, \rho)$.  The Cabibbo angle is denoted by $\theta = 0.22$.}
\begin{tabular}{|c|c|c|}
& $Im(\lambda_c/\lambda_t)$ & $Im(\lambda_c/\lambda_t)^2$ \\
\hline
$B_d$ & $\frac{\eta}{(1-\rho )^2 +\eta^2}$ & $\frac{-2\eta (1-\rho
)}{[(1-\rho )^2 +\eta^2 ]^2}$ \\
\hline
$B_s$ & $-\eta \theta^2$ & $2\eta \theta^2$
\end{tabular}
\end{table}

\end{document}